\documentclass[12pt]{iopart}

\usepackage{iopams}

\usepackage{graphicx}
\usepackage{harvard}

\usepackage{graphicx}
\usepackage{epstopdf, epsfig}

\usepackage{wrapfig}
\usepackage{lipsum}

\usepackage[utf8]{inputenc}
 
\usepackage{caption}

\usepackage{dblfloatfix}

\usepackage{dcolumn}
\usepackage{bm}
\usepackage{amsfonts}

\usepackage{color}

\usepackage{array}

\usepackage{setspace}

\usepackage{graphicx}
\usepackage{caption}
\usepackage{ragged2e}

\usepackage[font=scriptsize,labelfont=bf]{caption}

\usepackage{enumitem}

\usepackage{multirow}

\usepackage[superscript,biblabel]{cite}
\usepackage{textcomp}

\begin{document}

\title[Nozzle subtraction effects in sinonasal CFD modeling]{On computational fluid dynamics models for sinonasal drug transport: relevance of nozzle subtraction and nasal vestibular dilation}

\author[Basu, Frank-Ito, Kimbell]{Saikat Basu$^1$\footnote[1]{\scriptsize{Corresponding author. E-mail: saikat$\_$basu@med.unc.edu}}, Dennis O. Frank-Ito$^{2,3,4}$  and Julia S. Kimbell$^1$}

\address{$^1$Computing and Clinical Research Lab, Department of Otolaryngology/Head and Neck Surgery, The University of North Carolina -- School of Medicine, Chapel Hill, NC 27599, USA\\
                $^2$Div. of Head \& Neck Surgery \& Communication Sciences, Duke University Medical Center, Durham, NC 27710, USA\\
	     $^3$Computational Biology and Bioinformatics Program, Duke University, Durham, NC 27708, USA\\
               $^4$Department of Mechanical Engineering and Materials Science, Duke University, Durham, NC 27708, USA}

\begin{abstract}
Generating anatomically realistic three-dimensional (3D) models of the human sinonasal cavity for numerical investigations of sprayed drug transport presents a host of methodological ambiguities. For example, subject-specific radiographic images used for 3D reconstructions typically exclude spray bottles. Subtracting a bottle contour from the 3D airspace and dilating the anterior nasal vestibule for nozzle placement augment the complexity of model-building. So, we explored the question: how essential are these steps to adequately simulate nasal airflow and identify the optimal delivery conditions for intranasal sprays? In particular, we focused on particle deposition patterns in the maxillary sinus, a critical target site for chronic rhinosinusitis (CRS). The models were reconstructed from post-surgery computed tomography scans for a 39-year-old Caucasian male, with CRS history. Inspiratory airflow patterns during resting breathing are reliably tracked through CFD-based steady state laminar-viscous modeling and such regimes portray relative lack of sensitivity to inlet perturbations. Consequently, we hypothesized that the posterior airflow transport and the particle deposition trends should not be radically affected by the nozzle subtraction and vestibular dilation. The study involved 1 base model and 2 derived models; the latter two with nozzle contours (two different orientations) subtracted from the dilated anterior segment of the left vestibule. We analyzed spray transport in the left maxillary sinus for multiple release conditions. Similar release points, localized on an approximately 2mm-by-4.5mm contour, facilitated improved maxillary deposition in all three test cases. This suggests functional redundancy of nozzle insertion in a 3D numerical model for identifying the optimal spray release locations.
\end{abstract}

%



\section{Introduction}

Therapeutics for chronic rhinosinusitis (CRS) constitutes a layered approach\cite{fokkens2012, benninger2003, rosenfeld2007, albu2010} comprising  oral antibiotics, anti-inflammatory drugs, and surgical intervention, along with topical medications like nasal sprays. The anatomical alteration through surgery does not always address the inflammation from CRS, while long-term use of antibiotics and anti-inflammatory medications comes with systemic side-effects. Consequently, topical nasal sprays may represent a viable component of medical therapy for CRS. There is, however, a caveat as the topical sprays do not always ensure an optimum drug delivery to the affected areas of the sinonasal cavity. Of interest is hence to identify the different spray parameters and application techniques that would maximize the topical deposition of the drugs in critical areas like the maxillary and the ethmoid sinuses (see Figure~{\ref{f:coronal} for these anatomical landmarks). The techniques entail the head positions, nozzle positions, and breathing methods that would maximize the target site particle deposition (TSPD). While \textit{in vivo} measurements of drug delivery is an ambitious idea, computational fluid dynamics (CFD) simulations of nasal spray intake promise a feasible preliminary path towards identification of the optimal spray techniques.

\begin{figure*}
\includegraphics[width=14.2cm]{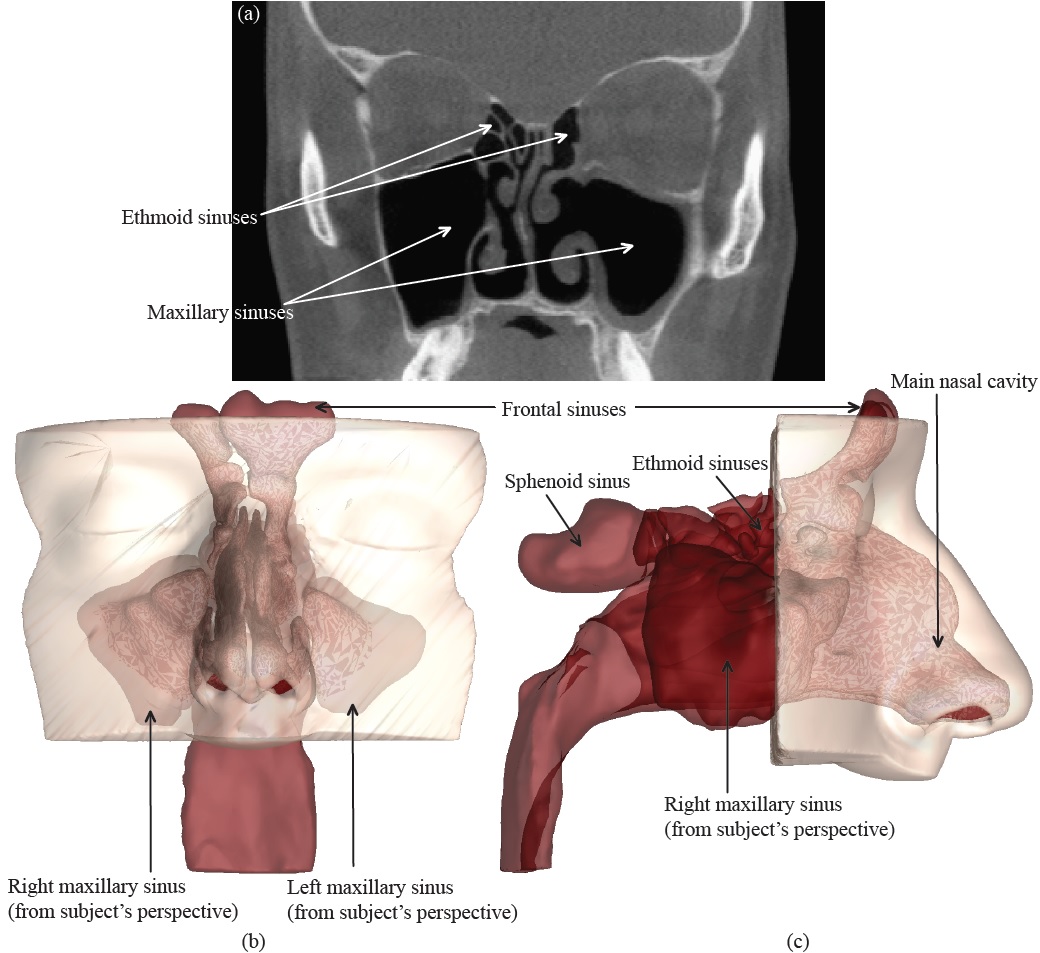}
\captionsetup{justification=justified}
\caption{(a) A representative CT imaging slice of the subject's sinonasal cavity, after a bilateral functional endoscopic sinus surgery. The section is anterior to the iatrogenic zone. Panel (b) depicts the coronal view and panel (c)  presents the sagittal view of the anatomically realistic 3D reconstruction of the sinonasal cavity from the CT scans. The different sinonasal chambers are marked.}\label{f:coronal}
\end{figure*}

Investigations through the CFD route incorporate the following three basic steps: (1) acquiring the patient-specific computed tomography (CT) scans of the sinonasal cavity, (2) developing an \textit{in silico} three-dimensional (3D) sinonasal model based on the CT scans using an image-processing software package, and creating a mesh of the flow domain for numerical simulations, (3) CFD analysis of the meshed model using numerical discretization techniques. Step (3) outputs the topical deposition of the medicated particles by tracking the flow patterns and the particle trajectories. To run a reliable simulation, it might seem essential to insert a nozzle into the 3D model in step (2) and edit the model by subtracting the nozzle contour from the nasal airspace, along with dilating the lateral walls to account for nozzle placement. The tip of the nozzle would indicate the starting point of the nasal spray particle transport. However, the patient-specific scans typically exclude any nozzle, and insertion of a nozzle at the image-processing stage involves an elaborate superficial reshaping of the anterior nasal cavity lining. There is no conclusive evidence in literature$^\mathrm{e.g.}\,$\cite{frank2012ohns, kimbell2007jas, shi2006jbe, inthavong2011jas, rygg2016jampdd} on the variability of the CFD-based modeling strategy to identify the optimal nasal spray instructions, based on whether the spray nozzle is accounted for or not in the CT-based 3D reconstructions. However, a preliminary study\cite{kimbell2015isam} in one normal subject suggested that the regional particle deposits were largely unaffected by the nozzle presence.

To address the above ambiguity related to the 3D modeling protocol, our study presents a comparison of the TSPD trends of nasal spray particles in three different test models (1 base model and 2 additional derived models) developed from the CT scans of the same patient. The models differed only at the anterior part of the nasal cavity at the left nostril. Model I (base model; see Figure~\ref{f:models}(a)) did not include any nozzle, while Model II and Model III (see panels (b) and (c) in Figure~\ref{f:models}) included spray bottles with two very different nozzle orientations (subtracted from the internal airspace), while still making sure that the spray directions conformed to the clinical safety guidelines. 

Nasal airflow for resting breathing has been evidenced to be predominatly laminar\cite{xi2008, shanley2008, kelly2000}, implying that the flow features are less sensitive to the inlet-zone perturbations resulting from the nozzle insertion in the anterior nasal vestibule. Thus, we hypothesized that the posterior airflow transport and consequently the sprayed particle depostion patterns in the three models should be relatable, which would in turn suggest redundacy of nozzle insertion in the numerical models while trying to identify the optimal spray release conditions. To check this hypothesis, we conducted particle simulations for a wide array of release points in the left nasal airspace and compared the locations for improved topical deposition in the left maxillary sinus (LMS) across the three test models. 

Preliminary reports on this work have been presented at the annual meetings of the American Physical Society (APS) -- Division of Fluid Dynamics\cite{basu2016dfd, basu2017dfd}.


\section{Methods}

\begin{figure*}
\centering
\includegraphics[width=14.2cm]{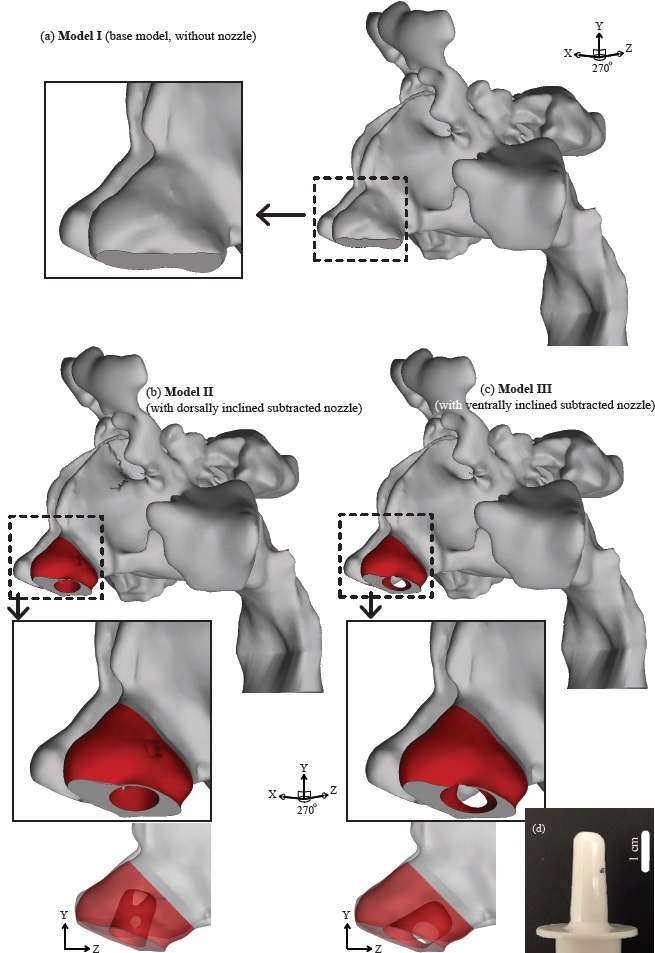}
\captionsetup{justification=justified}
\caption{(a) Model I (base model) reconstructed from the CT scans. Panels (b) and (c) show the derived models (Model II and Model III, respectively). In (b) and (c), the red contours comprise the nozzle-subtracted airspace and represent two different nozzle orientations. Panel (d) demonstrates the real nozzle contour that was subtracted from the base model airspace to generate Models II and III, each with a different angular orientation of the nozzle.}\label{f:models}
\end{figure*}

\begin{figure*}
\centering
\includegraphics[width=14.2cm]{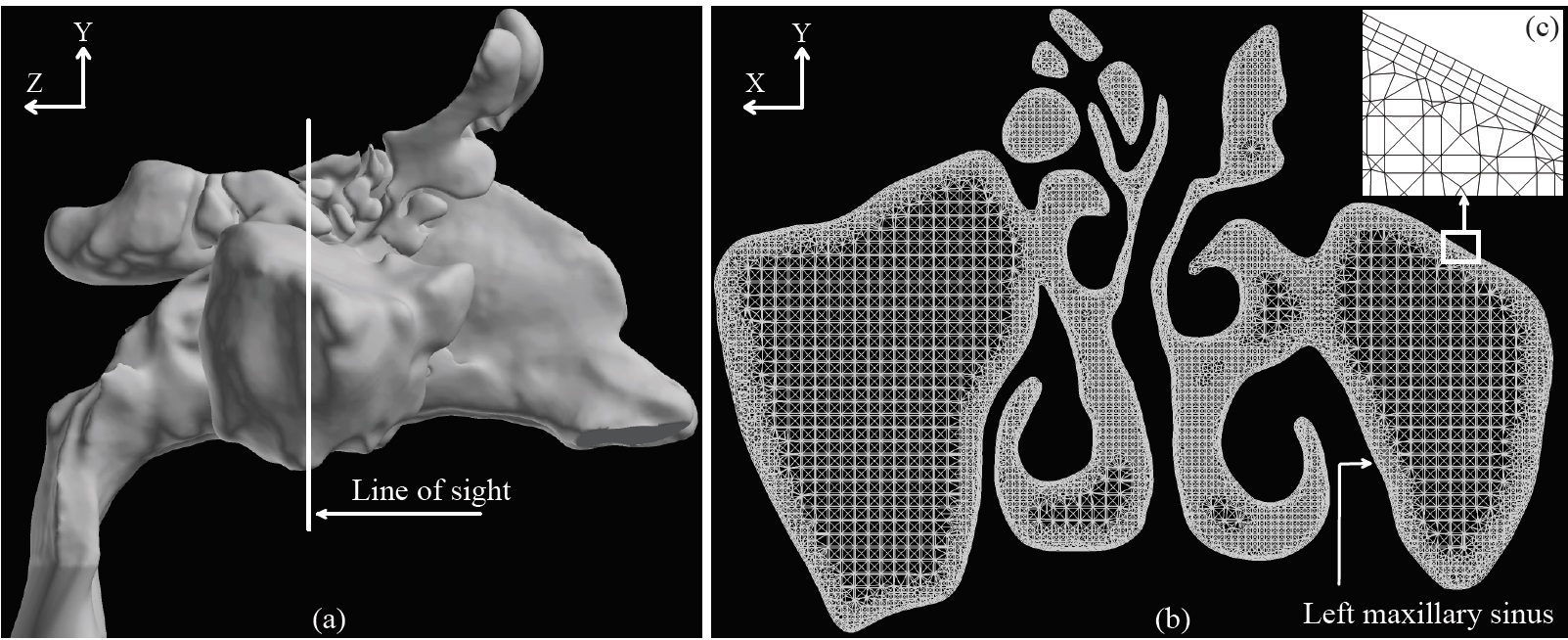}
\captionsetup{justification=justified}
\caption{(a) The white line on the sinonasal base model (Model I) illustrates a representative cross-sectional cut. (b) Representative visual of the meshed model along the cut from panel (a). The mesh consists of 4 million unstructured, graded tetrahedral elements along with three layers of 0.1 mm prism cells extruded at the cavity-tissue interfaces. (c) \textit{Inset}: A representative zoomed-in snapshot of the boundaries with the prism layers. Background color was inverted for better visualization. These figures were generated on the postprocessing software package FieldView\textsuperscript{TM} 16 (Intelligent Light, Lyndhurst, PA).}\label{f:meshing}
\end{figure*}

\subsection{Patient selection}
This methods study was implemented under the approval of the Institutional Review Board (IRB) committee at the University of North Carolina at Chapel Hill. It is part of an ongoing more extensive investigation\cite{dennis2016rdd} by our group to identify the optimal spray techniques and breathing methods for improved topical deposition of medicated aerosol particles, using the tools of CFD, supplemented by \textit{in vitro} experimental findings. We are targeting a final cohort recruitment of 30 CRS patients. The acquired CT scans, before and after functional endoscopic sinus surgery (FESS), would be used to develop 3D models for CFD-based explorations. In order to ensure a homogeneous study population, we only include patients without polyps and with the sinus disease confined primarily to the ostiomeatal complex (OMC) region. For the current study, we have used de-identified CT scans from a 39-year-old Caucasian male (weighing 94.5 kg with height 72 inches), who had a bilateral FESS for the treatment of medically recalcitrant CRS, without nasal polyps.

\subsection{Generating the base sinonasal model (Model I)}\label{ss:modelprotocol}
Anatomically realistic 3D reconstruction of the sinonasal airspace was developed from the post-operative CT scan images imported into the medical imaging software package Mimics\textsuperscript{TM} 18.0 (Materialize, Inc., Plymouth, MI, USA) in Digital Imaging and Communications in Medicine (DICOM) file format. The scans consisted of 270 slices, taken at depth increments of 0.399 mm, with a pixel size of 0.4 mm. An image radiodensity threshold range of -1024 to -300 Hounsfield units\cite{at2016ijnmbe} was used to delineate the nasal airways and the paranasal sinuses from the medical-grade CT scans, followed by careful hand-editing of the selected pixels to achieve anatomic accuracy. The sinonasal geometries were exported in stereolithography (STL) file format to the computer-aided design and meshing software ICEM-CFD\textsuperscript{TM} 15.0 (ANSYS, Inc., Canonsburg, PA, USA), where the 3D reconstruction was finalized in an $\mathrm{XYZ}$ cartesian space. For convenience, all the geometries were re-oriented in ICEM\textsuperscript{TM} to render the nasal floor parallel to the Z-axis. For the airspace boundaries: (1) planar surfaces were added spanning the outer rims of the nostrils, (2) a 2-cm outlet tube was added to the posterior end of the nasopharynx to ensure a fully-developed outlet flow and was closed off by a planar surface. The airspace was meshed using approximately four million unstructured, graded tetrahedral elements with three layers of 0.1 mm prism elements extruded at the air-tissue boundary. We performed a mesh quality analysis to minimize the number of  distorted low-quality elements, which could hinder the accuracy of the numerical simulations. Furthermore, as evidenced in literature\cite{frank2016influence} through a sensitivity analysis on the average pressure at the posterior septum and on the outlet flow, four million graded tetrahedral elements in the base computational mesh are sufficient to achieve robust and grid-independent numerical solutions. To finalize the meshing protocol, we further referenced our recent work\cite{basu2017jampdd} which has suggested that localized mesh refinement (on top of the base tetrahedral mesh with four million elements) in the sinuses is not essential for these simulations, based on an aymptotic analysis of inward and outward airflow ratio across the antrostomy window. This sinonasal model constituted the base model, referred to as Model I in the present study.

\subsection{Creating additional sinonasal models -- Model II and Model III:}

A 3D printer was used to print Model I in two compartments: (1) the anterior compartment, made from a flexible material, comprising the external nares and nasal vestibule; and (2) the posterior compartment comprising the rest of the sinonasal airway and made from a rigid material. CT imaging of the flexible anterior compartment with a nasal spray bottle inserted through the left nostril was used to generate a 3D rendition of the assembly in Mimics\textsuperscript{TM}. The spray bottle reconstruction was then virtually repositioned by aiming it toward the back of the nose in accordance to the package guidelines\cite{benninger2004techniques} for commonly prescribed nasal steroid spray products, and the nozzle was subsequently subtracted from the distended vestibular airspace. On the ICEM-CFD\textsuperscript{TM} platform, this distended anterior chunk was separated from the new model and replaced into a copy of Model I, from which the same left anterior undistended region had already been taken out. We developed two such derived models: Model II with an upward inclined spray axis and Model III with a downward inclined spray axis. Lowest panels in Figures~\ref{f:models} (b) and (c) provide visual distinction between the two contrasting nozzle orientations, when seen from the sagittal direction. These derived models were meshed following the same protocol as discussed above in Section~\ref{ss:modelprotocol}. Note that to maximize inter-model comparability, the models were kept identical except at the anterior half of the left nasal vestibule, where we had the nozzle insertion.

\subsection{Setting out the feasible release points for the nasal spray}\label{ss:releasepoints}

\begin{figure*}
\centering
\includegraphics[width=14.2cm]{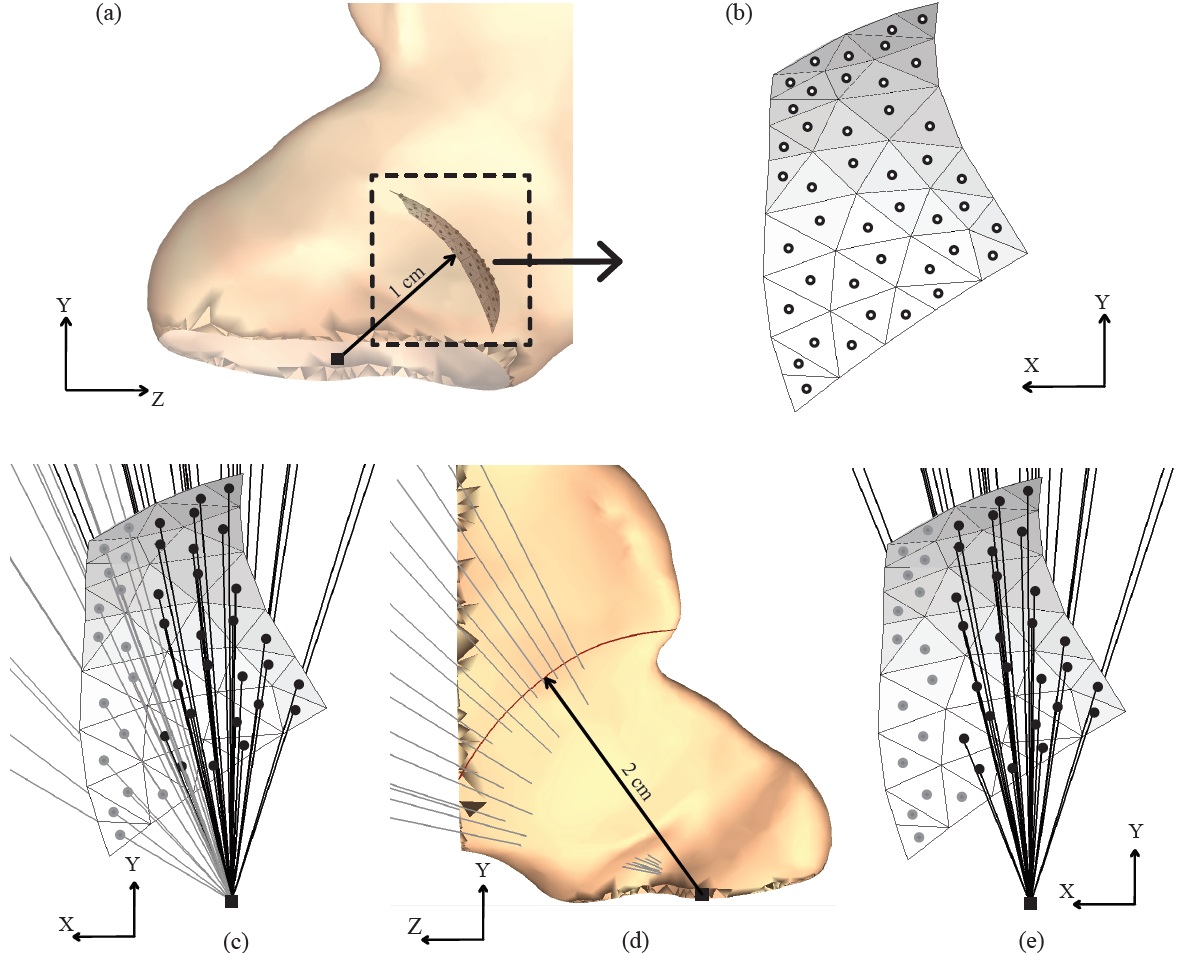}
\captionsetup{justification=justified}
\caption{Safe spray release points are extracted from Model I. The $\mathrm{X}$ axis points in the sagittal direction, the $\mathrm{Z}$ axis points in the coronal direction, and the $\mathrm{Y}$ axis points perpendicularly upward from the hard floor of the palate in the axial direction. Each individual spray release axis is directed from the centroid of the nostril plane (marked by the tiny dark square) to the corresponding release point. Panels (a) and (b) show the release contour. Panel (c) demonstrates all 43 probable release points and directions. Panel (d) sagittally lays out the clinically inadvisable directions. The 16 clinically safe release conditions are in panel (e).}\label{f:releasepoints}
\vspace{-0.5cm}
\end{figure*}

Figure~\ref{f:releasepoints} lays out the lattice orientation of 43 potential spray release points inside the left nasal vestibule. The contour was so selected as to mark the physical limits for comfortable nozzle placement. These points were all equidistant from the centroid of the planar inlet surface that covered the left nostril in the 3D model and at a depth of 1 cm from it. The angular location of these points covered a swept region of 75$^\circ$ to 30$^\circ$ to the horizontal. For Model I, the direction vectors for spray axis were calculated from the centroid of the left nostril plane to the corresponding release point. Of the 43 probable spray release points identified, the points for which the release directions were hitting the septum in the base model (Model I),  within a depth of 2 cm from the centroid of the left nostril plane, were considered to be clinically ``unsafe''. This is reasoned based on the intranasal spray guidelines\cite{benninger2004techniques} which suggest pointing the spray nozzle in the direction of the lateral nasal wall and away from the septum. The criterion precluded 16 release points in the current lattice from further consideration, and particle tracking simulations were performed using the remaining 27 ``safe'' spray directions. Note that in Models II and III, each spray axis was directed from the centroid of the area-outline on the left nostril plane where the nozzle contour intersects it, to the corresponding release point.

\subsection{Characteristics of the topical nasal spray}\label{ss:sprayfeatures}

For a realistic nasal spray transport and deposition in our CFD simulations, over-the-counter Flonase\textsuperscript{TM} Allergy Relief was selected as the spray to be numerically tracked for this study; it being one of the most commonly prescribed nasal sprays. Four units of Flonase\textsuperscript{TM} were sent to Next Breath, LLC (Baltimore, MD, USA) to test the \textit{in vitro} spray performance. The findings gave us the droplet size distribution as measured by laser diffraction using a Malvern Spraytec STP2000 and plume geometry as analysed by a SprayVIEW NOSP, which is a non-impaction laser sheet-based instrument. The mean spray half cone angle was observed to be 31.65$^\circ$, and the droplet sizes were found to follow a log-normal distribution. The probability density function, with $x$ being the droplet diameter, was of the form:
\begin{equation}
f(x) = \frac{1}{\sqrt{2\pi}x \ln \sigma_{g}} \exp \left[ -\frac{(\ln x - \ln x_{50})^2}{2 (\ln \sigma_g)^2}  \right].
\end{equation}
Here $x_{50} = 37.16$ $\mu$m is the volume mean diameter, or in other words, $\exp (x_{50})$ is the median of the lognormal distribution \cite{mould1998}. Also, $\sigma_g =2.08$ is the geometric diameter (geometric standard deviation), or in other words, $\ln \sigma_g$ is the standard deviation $S$, with $S^2 = \sum_{i=1}^{n}[\ln (x_i - \bar{x})]^2/(n-1)$, where $n$ is the total number of particulate droplets and $\bar{x} = \left( \sum_{i=1}^{n}\ln x_i\right)/n$. The distribution of the particulates with respect to their diameters is shown in Figure~\ref{f:psd}. With droplet sizes ranging from 5 to 525 $\mu$m in aerodynamic diameter (with size bin increments of 5 $\mu$m) and assuming them to be spherical and of unit density, the total number of droplets in one shot weight of $W_S=100$ mg quantified to 343,968. While simulating the particle trajectories in the CFD software package Fluent\textsuperscript{TM} 14.5 (ANSYS, Inc., Canonsburg, PA, USA), we assumed solid-cone type injections. Furthermore, for the mean spray exit velocity, we used 19.2 m/s based on phase doppler anemometry-based measurements in literature \cite{liu2011} for Flonase\textsuperscript{TM}.

\begin{figure*}
\centering
\includegraphics[width=14.2cm]{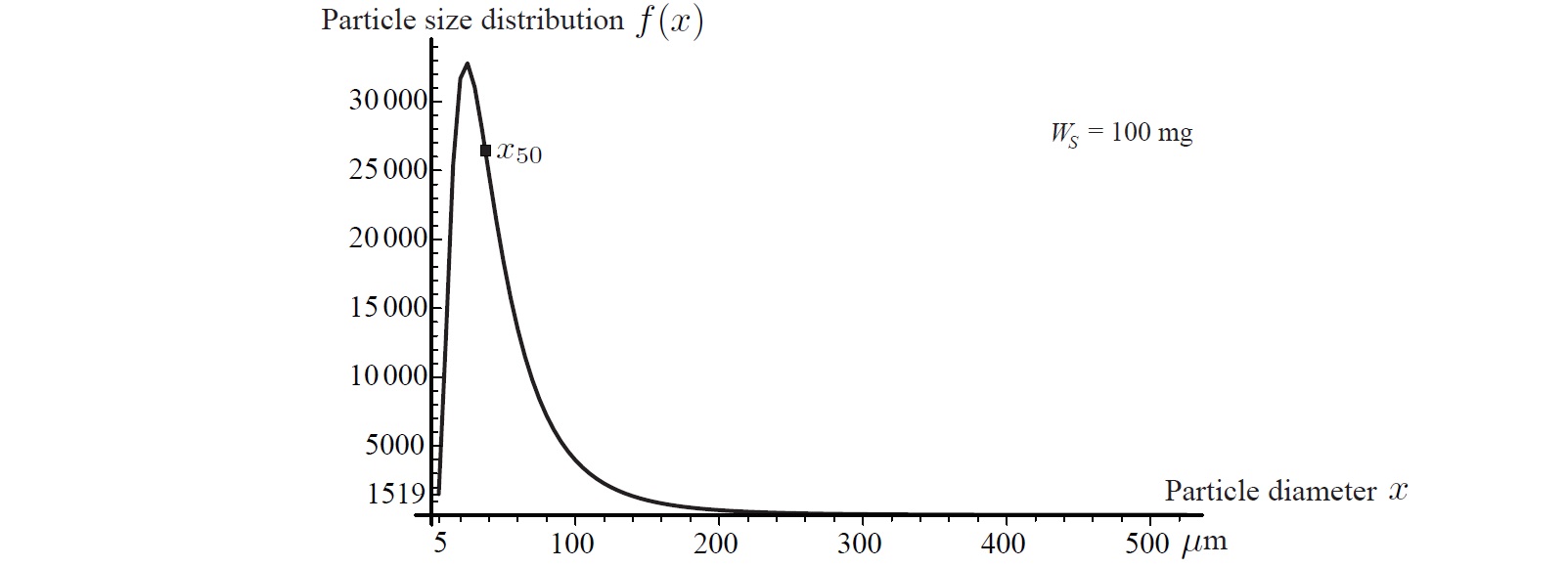}
\vspace{-2mm}
\captionsetup{justification=justified}
\caption{Lognormal distribution of the particulate droplet sizes in the Flonase\textsuperscript{TM} spray for a single shot weight of 100 mg. The total number of particles in one shot, as per the particle size distribution (PSD) with 5 $\mu$m size bins, was estimated to be 343,968. The particles were all assumed to be spherical in shape and of unit density. $W_S$ represents one shot weight.}\label{f:psd}
\end{figure*}

\subsection{Solution schemes for the inspiratory flow dynamics and the nasal spray trajectories}\label{s:mathematics}
For an incompressible fluid, the principle of mass conservation (\emph{continuity}) necessitates 
\begin{equation}\label{e:continuity}
\nabla \cdot \mathbf{u} = 0,
\end{equation}
with $\mathbf{u}$ being the fluid velocity vector. At steady state, the conservation of linear momentum leads to the modified Navier-Stokes equations:
\begin{equation}\label{e:NS}
\rho\left(\mathbf{u} \cdot \nabla \right)\mathbf{u} = -\nabla p + \mu {\nabla}^2 \mathbf{u}+\rho\mathbf{b},
\end{equation}
where, in the current system, $\rho = 1.204$ kg/m$^3$ is the air density, $\mu = 1.825\times10^{-5}$ kg/m.s is the dynamic viscosity of air, $p$ is the incumbent pressure, and $\mathbf{b}$ represents the body accelerations like gravity ($g$), inertial accelerations etc. Airflow simulations were performed by numerically solving the differential equations \ref{e:continuity} and \ref{e:NS} using Fluent\textsuperscript{TM} through a finite volume approach, under laminar conditions in the inspiratory direction. Justification for assuming flow-laminarity emerged from experimental evidences \cite{kelly2000} which suggested that nasal airflow belongs to the laminar regime at comfortable resting breathing rates. Also, since our simulation explored a single cycle of inspiration, steady state flow conditions were surmised to be a good approximation for resting breathing. The numerical solution employed the following boundary conditions: (1) zero velocity at the air-tissue interfaces (\emph{no slip} at the walls, with ``trap'' boundary conditions for the particle tracking); (2) zero pressure at the nostril planes which acted as the pressure-inlet zones (with ``reflect'' boundary conditions for particle tracking); and (3) a negative outlet pressure (with ``escape'' boundary conditions for particle tracking) commensurate to the inhalation airflow rate based on the subject-specific allometric scaling. Such equations, $\dot{V}=1.36\,M^{0.44}$ for males (sitting awake) and  $\dot{V}=1.89\,M^{0.32}$ for females (sitting awake), have been derived in published literature\cite{garcia2009} for a healthy cohort, with $\dot{V}$ being the minute volume in liters per minute and $M$ being the body mass in kilograms. The formulation (for male cohort) was judged applicable for the current post-operative subject owing to an assumed similarity of breathing patterns between non-symptomatic healthy people and post-operative CRS patients without nasal polyps. The solution scheme employed a segregated solver, based on SIMPLEC pressure-velocity coupling and second-order upwind spatial discretization. Numerical convergence was determined through minimizing the residuals (mass continuity $\sim10^{-2}$, velocity components $\sim10^{-4}$), and by stabilizing the mass flow rate and the static pressure at the outlet.

Particle transport was simulated through the series of solid-cone injections emanating from the laid out release points (see Section~\ref{ss:releasepoints}), with the spray characteristics as discussed in Section~\ref{ss:sprayfeatures}. Separate particle tracking simulations were performed for each of the 27 safe release points and the respective TSPD at the LMS were compared to identify the optimal release zones. The simulations were based on a Discrete Phase Model (DPM) in Fluent\textsuperscript{TM}, in which Lagrangian particle tracking was used to estimate the individual trajectories by integrating the particle transport equations through the Runge-Kutta method:
\begin{equation}
\frac{du_p}{dt} = F_D (u-u_p)+g\,\frac{\rho_p - \rho}{\rho_p} + F_i,
\end{equation}
with $u_p$ being the particle velocity, $u$ being the airflow field velocity, $\rho_p$ and $\rho$ being the particle and air densities respectively, $g$ being the gravitational acceleration, and $F_i$ being the additional forces per unit particle mass, like Saffman lift force contribution owing to the lift exerted by a flow-shear field on small particles perpendicular to the direction of flow. $F_D (u-u_p)$ is the drag force per unit particle mass, with
\begin{equation}
F_D = \frac{18 \mu}{\rho_p d_p^2}\frac{C_D \mathbb{Re}}{24},
\end{equation}
where $\mu$ is the dynamic viscosity of air, $d_p$ is the particle diameter, $C_D$ is the drag coefficient on a smooth spherical particle\cite{morsi1972jfm}, and $\mathbb{Re}$ is the relative Reynolds number, calculated as $\mathbb{Re} = \rho d_p |u_p-u|/\mu$.

The solution scheme considered the particulate droplets to be large enough to ignore any possible effects of Brownian motion on their dynamics. Spray particle simulations were conducted five times for each release point in all three models, to account for the variability of the Fluent\textsuperscript{TM}-based DPM solver in assigning random initial directions to the sprayed particles in the solid-cone injections. The topical particle deposition numbers in the LMS were expressed in terms of the deposited mass fraction (DMF) percentage, which was calculated as 
\begin{equation}
\textrm{DMF} = \frac{W_{\textrm{LMS}}}{W_S}\times 100,
\end{equation}
where $W_{\textrm{LMS}}$ was the net weight of the particulate droplets deposited in the LMS in each simulation run. To gauge the statistical significance of the congruity of the TSPD patterns in the three test models, we ran Spearman's rank-order test on the particle deposits at the LMS in each model and examined the rank correlation between the three sets of paired groups drawn from the test models.

\begin{figure*}
\centering
\includegraphics[width=14.2cm]{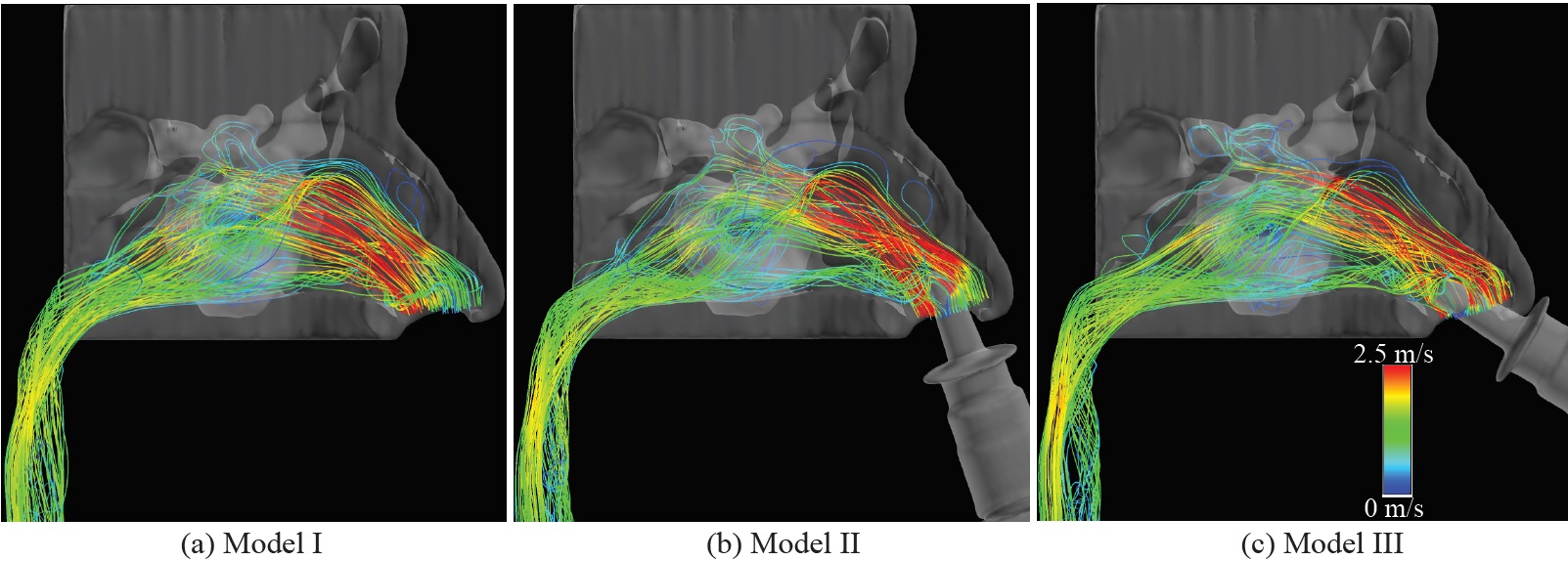}
\captionsetup{justification=justified}
\caption{Representative streamlines from the numerical airflow simulations in: (a)~Model I, (b)~Model II, and (c)~Model III. Panels (b) and (c) additionally demonstrate the nozzle placement, corresponding to the nozzle contours that have been subtracted (along with the inclusion of dilated nares) from the nasal airspace in the two derived models. The grey background shows the tissue domain lining the sinonasal cavity. These figures were generated using the postprocessing software package FieldView\textsuperscript{TM}, after import of the numerical solutions from Fluent\textsuperscript{TM}.}\label{f:flows}
\end{figure*}

\subsubsection{Monitoring possible regime transition in the nasal airflow:}
Vigorous air-intake during sniffing or higher respiratory demands during exercise can mobilize sinonasal flows to devolve into turbulence. However, at resting to moderate breathing rates ($\leq25$ L/min), more streamlined laminar airflow conditions are found to prevail\cite{xi2008, shanley2008, kelly2000}. For a consistent basis to identify the flow regime, we looked at Reynolds numbers in the three models. The dimensionless Reynolds number ($Re$), which is a ratio of the convective inertia of the flow to its viscosity, can be calculated\cite{white2015fluid} as $Re = \rho\, v\, D_h/\mu$, where $\rho$ is the air density, $v$ is the airflow speed, $D_h=$ 4$\times$(cross-sectional area)/(wetted perimeter at the cross-section) is the hydraulic diameter for irregular cross-section, and $\mu$ is the dynamic viscosity of air. In the airflow simulations, we measured $Re$ at the cross-section where the anterior segment from Model I was replaced by the nozzle-subtracted contours to generate Models II and III (see Figure~\ref{f:models}). The nasal cavity widened beyond that region, resulting in the $Re$ locally peaking close to that cross-section; it being directly proportional to the flow-rate which is, in turn, inversely related to the cross-sectional passage area transverse to the generic flow direction. Such a threshold choice helped us to monitor any possible transition to turbulent behavior as the $Re$ peaked.


\section{Results}

\subsection{Topical deposition of the nasal spray}



\begin{table}[t]
 \caption{Steady state inpiratory flow rates  (= 2$\times$minute volume) in Liters/min in the numerical simulations.}\label{table1}
\begin{center}
    \begin{tabular}{cccc}
    \hline
 \scriptsize{Allometric target} & \scriptsize{Model I} & \scriptsize{Model II} & \scriptsize{Model III} \\ \hline

   \scriptsize{20.13} & \scriptsize{20.0} & \scriptsize{20.17} & \scriptsize{20.15}\\
  \scriptsize{Variation from target} & \scriptsize{$-0.65$\%} & \scriptsize{$+0.20$\%} & \scriptsize{$+0.10$\%}\\
    \hline
    \end{tabular}
\end{center}
\end{table}

Effectivity of a topical medication to treat CRS critically centers on whether the sprayed particles can penetrate through to the typically remote sinonasal chambers, like the maxillary or the ethmoid sinuses. To compare the airflow and particle simulation results in the three models: we representatively collated the TSPD data in the LMS for the spray that was administered through the left nasal airspace in all the models. With respect to an allometric target of 20.13 Liters/min (see Section~\ref{s:mathematics} for the predictive formulation; also see Table~\ref{table1}), the laminar steady state flow simulations were converged at $\leq |1\%|$ error and returned inspiratory breathing rates of 20.0 Liters/min in Model I, 20.17 Liters/min in Model II, and 20.15 Liters/min in Model III.

\begin{figure*}
\centering
\includegraphics[width=14.2cm]{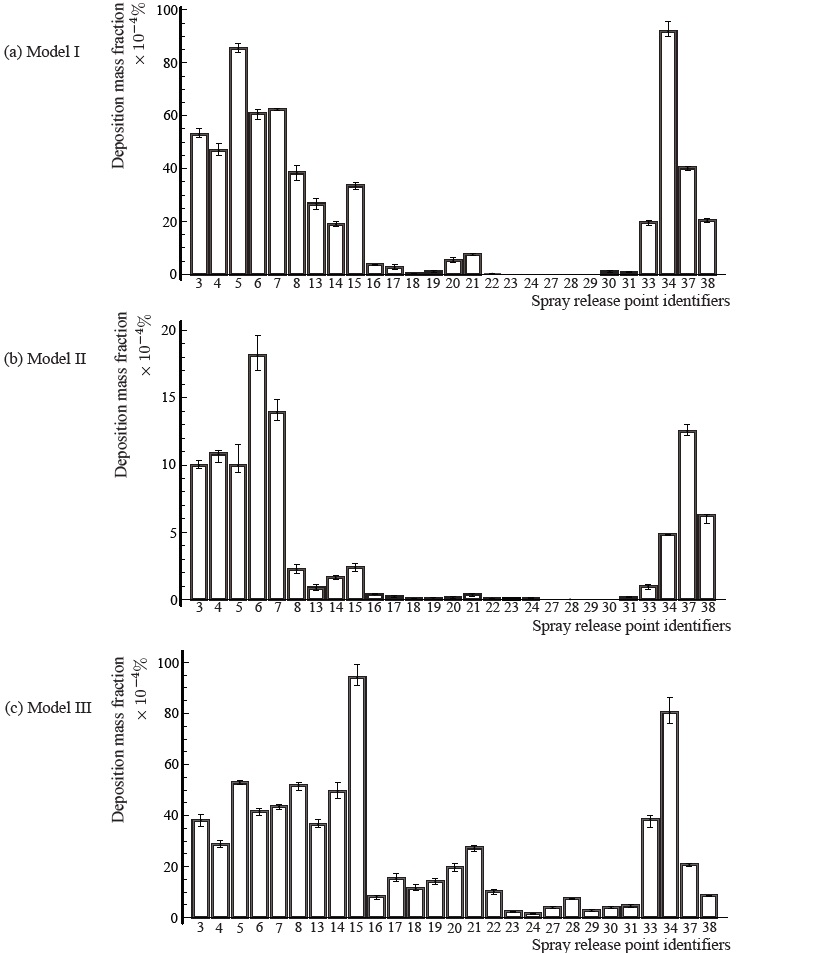}
\captionsetup{justification=justified}
\caption{Bar diagram of the spray mass fractions deposited in the LMS in the three test models. The numbers on the horizontal axis indicate the corresponding identifiers of the spray release locations, based on their assigned nomenclature in Fluent\textsuperscript{TM}. The same number in all the models corresponds to the same release location. The error bars represent the maximum and minimum deposits from the five particle tracking simulation runs implemented for each spray release point.}\label{f:depositions1}
\vspace{-0.5cm}
\end{figure*}

\begin{figure*}
\centering
\includegraphics[width=14.2cm]{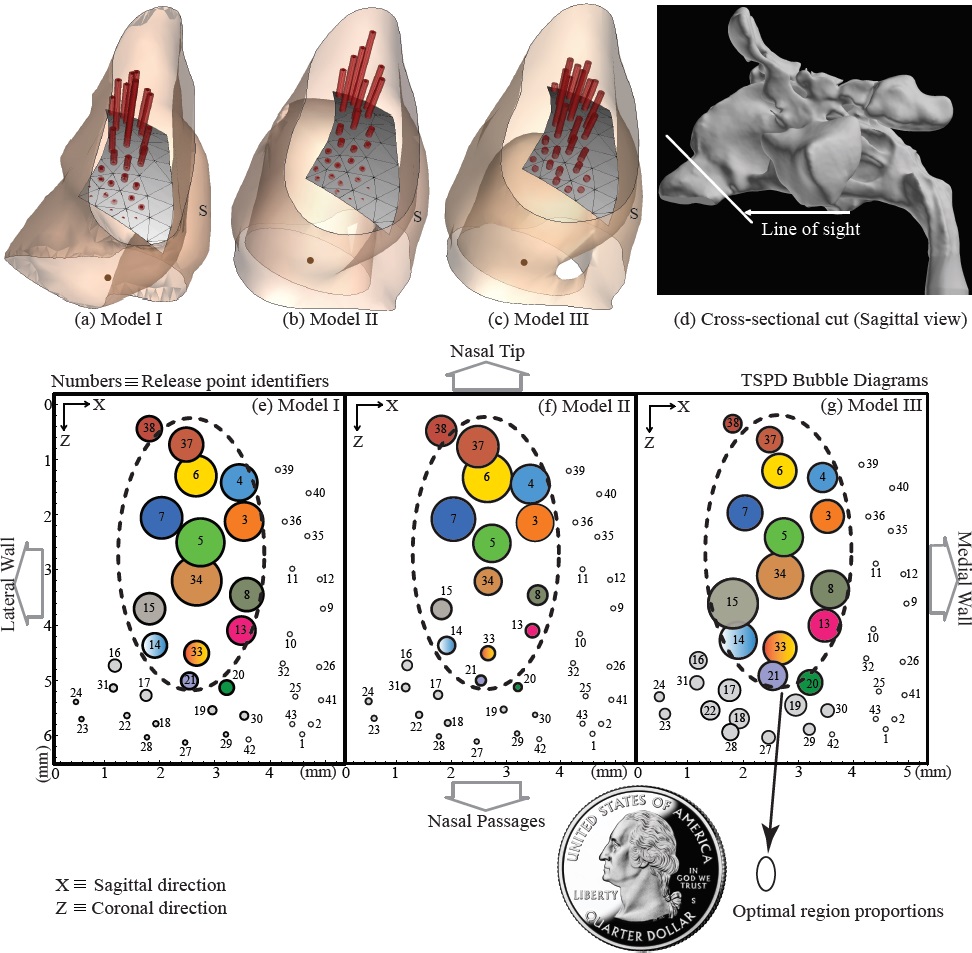}
\captionsetup{justification=justified}
\caption{Panels (a), (b), and (c) show the topical deposition patterns at the LMS in Models I, II, and III respectively. The height of the red spike at each release point is proportional to the deposited mass fraction corresponding to that spray release location. The spike heights have been proportionally standardized with the peak deposit in each model being represented by a 5 mm tall spike. Direction vectors of the spikes are oriented from the centroid of the nostril (in Model I) and from the centroid of the nozzle base on the nostril plane (in Models II and III) to the corresponding release point. These centroids are illustrated in the three pictorial representations by a small dark circle. The letter ``S'' marks the septal side. Panel (d) shows the sectional cut through which the depictions in (a), (b), and (c) were visualized. The ``bubble-diagrams'' in panels (e), (f), and (g) show an alternate visual representation of the same information. Here the release contour has been projected on the $\mathrm{XZ}$ plane, roughly parallel to the floor of the palate. The bubble sizes are proportional to the LMS deposit corresponding to the release location whose projection is the center of that bubble. The color scheme was so chosen that the findings for the same release point have the same color, in all three models. Note the unsafe release points were marked by the tiny hollow circles. We identified a small optimal zone for the best-possible spray release points and the ellipse (drawn with the dashed line) roughly demarcated that zone. For scaled representation, its size is compared to the US quarter-dollar coin.}\label{f:patterns}
\vspace{-0.5cm}
\end{figure*}

\subsection{Justification for the laminar idealization based on Reynolds numbers}
The $Re$ in Model I was 1945, followed by 1991 and 1989 in Models II and III, respectively. The anterior nasal flow can be idealized as a pipe flow and there is a deluge of experimental as well as numerical work in the applied mechanics literature on the flow transition in such systems. Under constant pressure gradient\cite{wygnanski1973}, turbulent ``puffs'' appeared at $Re$ $\approx$ 2800, with slugs (trapped bubbles resulting in two-phase flow) emerging beyond $Re$ $\approx$ 3000. For constant flow rates\cite{darbyshire1995}, fully-developed turbulence was seen beyond $Re$ $\approx$ 2700. The airflow in our models can hence be assumed to adhere to a predominantly laminar profile, with some probable transitional features. In this context, we noted that while the flow patterns (see Figure~\ref{f:flows}) demonstrated vortex formations, there are enough evidences$^\mathrm{e.g.}\,$\cite{basu2017jfm, stremler2014fdr, stremler2011jfs, basu2015pof, basu2014thesis} suggesting that a flow with such rotational patterns can still preserve laminarity. As a numerical idealization, implementing the laminar modeling framework was thus deemed justified.

\begin{figure*}
\centering
\includegraphics[width=14.2cm]{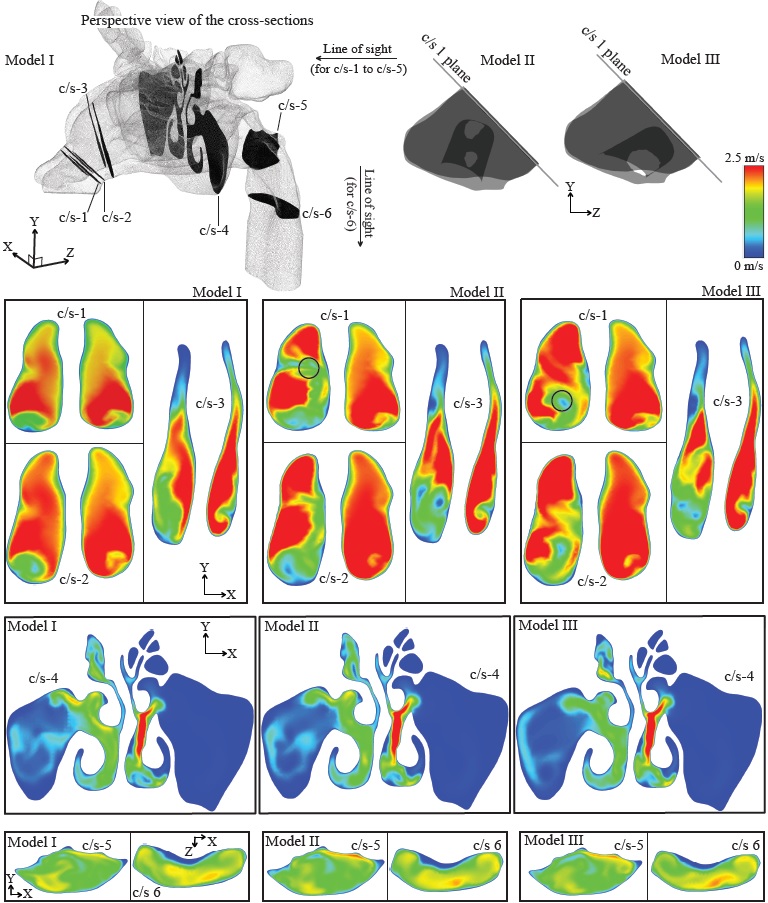}
\captionsetup{justification=justified}
\caption{Color maps of the velocity magnitude of the inspired airflow field across six different cross-sections (c/s-1 to c/s-6). See the top-most panel for the cross-sections selected. Left-right orientation of the graphics in each of the flow profile panels respectively corresponds to the left and right sides of the study subject. The flow patterns observed on c/s-1 noticeably demonstrate the fluctuations from nozzle placement (in Models II and III) in the anterior vestibule. Stagnation zones closely posterior to the nozzle tips are demarcated by the black circles. Flow profiles display increasing conformity at larger penetration depths.}
\end{figure*}

\subsection{Congruity of deposition patterns in the three models}\label{congruity}
Figure~\ref{f:depositions1} plots the mass fraction of spray particles deposited in the LMS for each of the 27 safe release points (see Section~\ref{ss:releasepoints} for details on the spray release points and release contour). Based on the numerical identity of the release points (along the horizontal axis in Figure~\ref{f:depositions1}), the deposition patterns, in terms of the peak and plateau zones, were congruous for the three  models, with the optimal release points still localized in the same areas of the release contour. In this context, Figures~\ref{f:patterns}(a)--(c) demonstrate an alternate view with the extruded spikes at each release point denoting the quantity of LMS deposits as well as the direction vector of the original spray. In Figures~\ref{f:patterns}(e)--(g), we projected the release contour on to the $\mathrm{XZ}$ plane (parallel to the floor of the hard palate) and traced a ``bubble-diagram'' with the incrementing circle sizes representing proportionally higher topical deposits at the LMS, with the center of each circle being the projection of the corresponding release point. Note that the color schematic in Figures~\ref{f:patterns}(e)--(g) is such that the same release point has the same color in the three planar projections. A small optimal zone for the best-possible spray release points was identified and it was the same in all the test models. An ellipse (drawn with the dashed line), spanning roughly over a 2mm-by-4.5mm contour, was drawn to demarcate that zone in the ``bubble-diagrams''.

For a statistical check of the congruity of TSPD patterns with respect to the release locations, we ran the Spearman's rank correlation test over three different selections of paired groups (Models I and II, Models I and III, and Models II and III) with respect to the corresponding LMS deposits. For each model, the test ranked the deposits from the different release points according to their magnitudes  and then assessed how well these rank-orders correlate between two different models. For the comparison between Models I and II, the test returned $R = 0.929$, with the two-tailed $p<0.000001$.  For Models I and III, we had $R = 0.866$, with the two-tailed $p<0.000001$.  For Models II and III, the test gave $R = 0.767$, with the two-tailed $p< 0.000003$. Thus, the congruity of deposition patterns across the three test models could be considered statistically significant.

\begin{figure*}
\centering
\includegraphics[width=14.2cm]{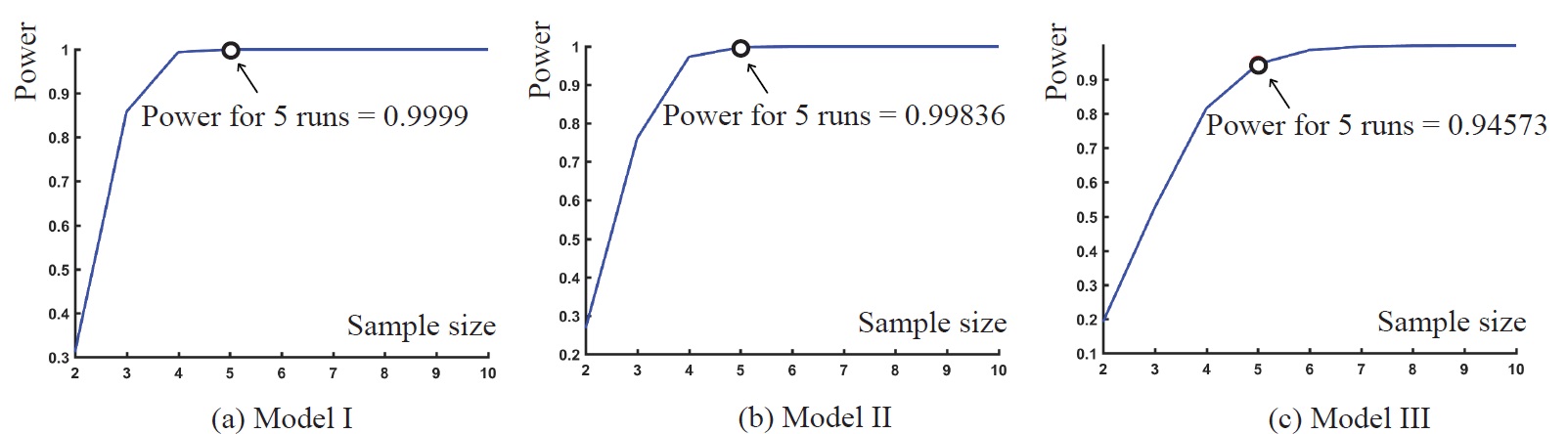}
\captionsetup{justification=justified}
\caption{Graphical representation of the power of the sample with respect to the sample size. The latter, in this case, is the number of numerical runs implemented for particle tracking corresponding to each release point. These representative power curves were traced out for the release location 1 in all the three test models.}\label{f:power}
\end{figure*}

\begin{table}[b]
\caption{Mean mass fraction deposits at the left maxillary sinus, based on 5 runs, along with the corresponding standard deviations. Left-most column indicates the spray release point identifiers.}\label{table2}
\begin{center}
\resizebox{\textwidth}{!}{%
\begin{tabular}{@{}|c|c|c|c|c|c|c|@{}}
\hline
\multirow{2}{*}{\begin{tabular}[c]{@{}c@{}}Rel.\hspace{-1cm}\\ Pt. \hspace{-1cm}\end{tabular}} & \multicolumn{2}{c|}{Model I}                                 & \multicolumn{2}{c|}{Model II}                                                     & \multicolumn{2}{c|}{Model III}                                                    \\ \hline
                                                                                    & \multicolumn{1}{l|}{Mass fraction} & \multicolumn{1}{l|}{Standard deviation} & \multicolumn{1}{l|}{Mass fraction} & \multicolumn{1}{l|}{Standard deviation} & \multicolumn{1}{l|}{Mass fraction} & \multicolumn{1}{l|}{Standard deviation} \\ \hline

3                                                                                   & 0.00529            & 0.000127279                             & 0.0009984                               & 0.0000294669                            & 0.003818                                & 0.000216032                             \\
4                                                                                   & 0.004678           & 0.000176833                             & 0.001082                                & 0.0000363318                            & 0.00288                                 & 0.000112472                             \\
5                                                                                   & 0.008526           & 0.000140285                             & 0.0009966                               & 0.0000875717                            & 0.005278                                & 0.0000834865                            \\
6                                                                                   & 0.006098           & 0.000110995                             & 0.001812                                & 0.000107564                             & 0.004156                                & 0.0000988939                            \\
7                                                                                   & 0.006236           & 0.0000328634                            & 0.001392                                & 0.0000641872                            & 0.004344                                & 0.0000867756                            \\
8                                                                                   & 0.003834           & 0.000281034                             & 0.000224                                & 0.0000260096                            & 0.005174                                & 0.000127201                             \\
13                                                                                  & 0.00266            & 0.000162635                             & 0.0000861                               & 0.0000190164                            & 0.003658                                & 0.000127554                             \\
14                                                                                  & 0.001896           & 0.0000673053                            & 0.0001612                               & 0.0000119038                            & 0.00496                                 & 0.000293002                             \\
15                                                                                  & 0.003342           & 0.000127945                             & 0.0002386                               & 0.0000203052                            & 0.009408                                & 0.000335738                             \\
16                                                                                  & 0.000357           & 0.0000175642                            & 0.0000379                               & 0.00000257196                           & 0.000831                                & 0.0000677237                            \\
17                                                                                  & 0.0002654          & 0.0000656415                            & 0.0000236                               & 0.00000651038                           & 0.00153                                 & 0.000121037                             \\
18                                                                                  & 0.00002524         & 0.0000133129                            & 0.000007196                             & 0.00000193154                           & 0.001154                                & 0.0000947629                            \\
19                                                                                  & 0.00010984         & 0.0000362876                            & 0.00000783                              & 0.00000432576                           & 0.001432                                & 0.0000952365                            \\
20                                                                                  & 0.0005118          & 0.0000802322                            & 0.000016644                             & 0.00000599314                           & 0.001952                                & 0.00014025                              \\
21                                                                                  & 0.0007564          & 0.0000371053                            & 0.000037                                & 0.00000955693                           & 0.002732                                & 0.0000960208                            \\
22                                                                                  & 0.000027158        & 0.0000131633                            & 0.000008688                             & 0.00000149655                           & 0.0010216                               & 0.0000763891                            \\
23                                                                                  & 0                  & --                                       & 0.00000485                              & 0.00000397665                           & 0.0002336                               & 0.0000124218                            \\
24                                                                                  & 0                  & --                                       & 0.000007704                             & 0.0000031044                            & 0.0001592                               & 0.0000117558                            \\
27                                                                                  & 0                  & --                                       & 0                                       & --                                       & 0.0004008                               & 0.0000281727                            \\
28                                                                                  & 0                  & --                                       & 0.00000183152                           & 0.0000013478                            & 0.0007478                               & 0.0000187003                            \\
29                                                                                  & 0                  & --                                       & 0                                       & --                                       & 0.0002794                               & 0.0000307782                            \\
30                                                                                  & 0.0001037          & 0.0000289418                            & 0                                       & --                                       & 0.0003976                               & 0.0000254224                            \\
31                                                                                  & 0.00007694         & 0.0000139414                            & 0.00001854                              & 0.00000485829                           & 0.0004614                               & 0.0000478571                            \\
33                                                                                  & 0.001952           & 0.0000903881                            & 0.00009594                              & 0.0000158248                            & 0.003846                                & 0.000199324                             \\
34                                                                                  & 0.009186           & 0.000258225                             & 0.0004806                               & 0.00000585662                           & 0.00803                                 & 0.000433359                             \\
37                                                                                  & 0.004022           & 0.0000491935                            & 0.001252                                & 0.0000334664                            & 0.002088                                & 0.0000593296                            \\
38                                                                                  & 0.002024           & 0.0000522494                            & 0.0006216                               & 0.0000299717                            & 0.000858                                & 0.0000256515 \\     
\hline                   
\end{tabular}}
\end{center}
\end{table}

\subsubsection{Varying penetration of spray particles in the three models:}\label{differences}
There was a caveat in terms of the absolute quantities of topical deposition. The peak deposit of 0.92\% spray mass fraction in Model I was more than five times the peak deposit (0.18\% spray mass fraction) in Model II; reason being larger anterior deposits in the latter. However, Model I and Model III presented somewhat similar peak deposits, with the latter offering  0.94\% spray mass fraction. The deposited particle sizes ranged from 5 to 40 $\mu$m.

\subsection{Variability of the numerical results for spray deposition}
The spray simulations were run 5 times for each release point, in each of the Models I, II, and III. We performed sample size analysis in each model to calculate the power of detecting a significant difference so that our mean deposition mass fraction based on the 5 simulation runs would lie within a margin (say, $\textrm{x}$) of error of 5\% of the peak mass deposition fraction in the LMS. Thus, for Model I, $\textrm{x}=0.00045$; for Model II, $\textrm{x}=0.00009$; and for Model III, $\textrm{x}=0.00047$. Simultaneously, the level of significance, which is the probability of rejecting the hypothesis when it is true (Type I error), was set at 0.05. With further inputs of the mean and the standard deviations of the deposited mass fractions; the power of the test, which quantifies the probability of correctly rejecting the hypothesis when it is false, came out to be very close to 1.0 in each model, when averaged over the spray release points. Specifically, the mean powers were 0.98 in Model I, 0.93 in Model II, and 0.96 in Model III. Hence, in terms of number of simulation trials at each release point, the study was considered detailed enough with a sufficiently broad sample size to ensure statistical reproducibility. This implementation scheme also conformed to previously published statistical benchmarking\cite{frank2012power}.


\section{Discussion}
We explored methodological ambiguities related to the inclusion of a spray bottle inside the anterior nasal vestibule of a 3D computational reconstruction, in the context of numerically ascertaining the optimal spray release locations for improved topical drug delivery. Three test models were developed: one base model without nozzle subtraction and two additional derived models with two very different orientations of subtracted nozzle from the nasal airspace (along with laterally dilated anterior vestibule to account for nozzle insertion), developed from the post-surgical CT scans of the same subject with a history of CRS. We performed inspiratory airflow simulations under steady state conditions in these three models. Sprayed particle transport was then numerically tracked for a lattice of 27 clinically safe nasal spray directions. The optimal zone for spray release for better drug penetration to the maxillary sinus (a crucial target site for treating CRS) stayed the same in all three models. The flow features were such that they could be reliably modeled within a laminar framework. Hence, the congruous TSPD patterns in the three cases supported our key hypothesis that laminar airflow regimes, being dictated by the geometric macro-scales, would be less sensitive to the inlet perturbations (pertaining to the nozzle placement zone) thereby hardly affecting the posterior airflow transport and particle deposition trends. The optimal spray release zone (see Figure~\ref{f:patterns}) was rather tiny and covered a roughly 2mm-by-4.5mm elliptical contour in all three test models. The results thus indicate a redundancy of nozzle insertion (i.e. subtracting the nozzle contour from the nasal vestibule) for such 3D anatomic reconstructions.

There are however some points of critique connected to the study design and assumptions, examined hereunder:
\begin{itemize}[leftmargin=*]
\item Reduction of the inspiratory airflow in resting human breathing to a steady state laminar profile, while being a good approximation for the brief single-cycle span during which the sprayed particles propagate through the sinonasal cavity, is still an idealization. Inspiration and expiration demonstrably have some acceleratory and deceleratory components. However, our preliminary forays and other existing findings\cite{chen2011ajra} on time-dependent simulations have until now yielded results in agreement with the steady state framework assumed for resting breathing.

\item The three models differed only in the anterior portion (refer back to Figure~\ref{f:models}) of the left nasal airspace. Model I was built from the subject-specific CT scans, while Models II and III had the anterior portions built from scans of a 3D printed model (of Model I) with a spray nozzle inserted at the left nostril. To compare between the models, care was taken that apart from this nozzle placement zone, the computational subject-specific reconstructions stayed analogous and comparable. However, note should be made that the subtle geometrical effects of the nozzle placement on the nasal lining may extend beyond this anterior chunk, based on subject-specific variations in tissue pliability.

\item Topical deposits at the (left) maxillary sinus were miniscule and less than 1\% spray mass fraction. The TSPD also differed between the three test models (for details, see Sections~\ref{congruity} and \ref{differences}). Such variations between numerical models with and without the spray nozzle are in agreement with published results\cite{inthavong2011jas}. However, the core aim of our study was to explore if nozzle subtraction and vestibular dilation were essential to identify the release points which would allow for the best possible sinonasal target site deposition. Thus, the absolute quantities of deposits were not of any compelling relevance and we focused more on the pattern of targeted deposition with respect to the lattice of spray release locations.

\item While quantifying the drug transport to the target sinonasal sites (in this study, the LMS), we have assumed that the sprayed particles get trapped on the internal cavity walls (mimicking the tissue surfaces), without any further mucus drainage transport or escape probabilities. This is an idealized assumption at this stage, and it would be interesting to capture the nuances beyond the scope of these current limitations. Nasal mucus is essentially non-Newtonian in terms of its physical properties and this makes modeling its viscosity challenging\cite{quraishi1998coas}. In this context, the reader may refer to a recent CFD-based study\cite{rygg2016jampdd} investigating the effects of mucociliary clearance on drug uptake in the human nasal cavity (excluding the paranasal sinuses). Also see\cite{wootton1999arbe} for a particularly interesting account (albeit not for respiratory flows) on the complex fluid dynamical subtleties associated with other classes of biological flows.

\item We assumed that the particles in motion never had any kinematic interactions with the other sprayed particles during their sinonasal transport. Nor was there any effect of the partcles' motion on the ambient inspiratory airflow. Additionally, their material density, sizes, and spherical shapes stayed unchanged as they propagated through the nasal cavity. 
\end{itemize}

Despite these caveats, this methods study, to summarize, does make the important suggestion that vestibular nozzle insertion in a 3D nasal model is not essential for reliable airflow simulations, particularly while identifying the optimal release conditions for targeted delivery of sprayed intranasal drugs. The inference alleviates the process of reconstructing anatomically realistic sinonasal models from patient-specific CT imaging. The findings will provide an important functional tool towards using 3D computational modeling on large cohorts of patient scans, which can help in recommending new usage instructions for aerosolized sprays aimed at maximizing particulate transport and deposition at the nasal target sites, hence providing a more effective therapeutic care for diseases like chronic rhinosinusitis.\\


\newpage

\ack \textbf{This is the pre-peer review version.}\\ Reported research was supported by the National Heart, Lung, and Blood Institute of the National Institutes of Health (NIH), under award number R01HL122154.  The content is solely the responsibility of the authors and does not necessarily represent the official views of the NIH. The authors thank the Rhinology Fellows and the Clinical Faculty at the Division of Rhinology, Allergy, and Endoscopic Skull Base Surgery (UNC School of Medicine -- Department of Otolaryngology/Head and Neck Surgery), in particular Adam M. Zanation, MD; Brian D. Thorp, MD; Charles S. Ebert Jr., MD; and Brent A. Senior, MD for their comments and suggestions, including feedback on model building and on identifying the clinically inadvisable nasal spray directions. Thanks are also due to Cara Breeden, BS and Julie D. Suman, PhD at Next Breath, LLC for furnishing the experimental findings on spray emission characteristics. Finally, the authors thank Nichole Witten, BS for technical help and many resourceful discussions.\\

\medskip

\begin{centering}
\scriptsize{CONFLICT OF INTEREST STATEMENT}\\
\end{centering}
\noindent The authors do not have any financial and personal relationships with other people or organization(s) that could inappropriately influence or bias their work. No conflict of interest exists in the submission of this manuscript, and the manuscript was approved by all the  authors for publication. The work described is original research that has not been published previously, and not under consideration for publication elsewhere. All the authors were fully involved in the study and preparation of the manuscript.\\

\medskip

\section*{References}
\bibliographystyle{wileyj}
\bibliography{ijnmbe2017}

\end{document}